\documentclass[review,authoryear]{elsarticle}
\usepackage{amsmath}%
\usepackage{amsfonts}%
\usepackage{amssymb}%
\usepackage{graphicx}

\begin{document}

\title{Decomposing data sets into skewness modes}

\author[knmi]{Rub\'{e}n A. Pasmanter\corref{cor1}}
\ead{R.A.Pasmanter@uva.nl}
\author[knmi]{Frank M. Selten}
\ead{selten@knmi.nl}
\cortext[cor1]{Corresponding author}
\address[knmi]{Royal Netherlands Meteorological Institute, POBox 201,3730AE, De
Bilt, Netherlands}

\begin{abstract}

We derive the nonlinear equations satisfied by the coefficients of linear
combinations that maximize their skewness when their variance is constrained
to take a specific value. In order to numerically solve these nonlinear
equations we develop a gradient-type flow that preserves the constraint. In
combination with the Karhunen-Lo\`{e}ve decomposition this leads to a set of
orthogonal modes with maximal skewness. For illustration purposes we apply
these techniques to atmospheric data; in this case the maximal-skewness modes
correspond to strongly localized atmospheric flows. We show how these ideas
can be extended, for example to maximal-flatness modes.
\end{abstract}
\maketitle


{\noindent {\bf Pacs:} 05.45.Tp, 92.60.Ry, 02.50.Sk}\\

\section{Introduction}

When dealing with large data sets it is convenient and costumary to make use
of the Karhunen-Lo\`{e}ve decomposition \citep{Karhunen1947} in order to express the data in terms
of the so-called empirical orthogonal functions (EOFs), 
these are linear
combinations of the original variables whose second-order cross-correlations
vanish, i.e. they are linearly uncorrelated, as in (\ref{covarDiago}). A
typical application consists in reducing then the number of degrees of freedom
to a relatively small number of EOFs with a variance larger than a certain
threshold, for more applications see for example \citep{Preisendorfer1988pca}. 
In many systems of interest the EOFs are nonlinearly correlated, a
fact that can have important consequences. Similarly, the probability
densities of the time-dependent EOFs' amplitudes are often approximately
Gaussian but the deviations from Gaussianity may be of great relevance. One
indicator of nonlinear correlations and of the non-Gaussian character of
fluctuations is the skewness, the third-order moments of the variables'
distribution. In this article we show how to construct orthogonal linear
combinations of \ the variables that maximize the skewness and we present a
numerical method in order to solve the ensuing nonlinear equations. For the
purpose of illustration we apply it to meteorological data; the
maximal-skewness modes found in this case correspond to spatially localized and
meteorologically meaningfull atmospheric flows.

\section{Maximal skewness modes}

Given a set of $n$ dynamical variables,%
\[
v_{i}(t),\ \ \ \ \ 1\leq i\leq n,
\]
with vanishing mean we want to construct a linear combination $\psi(t)$%
\[
\psi(t):=\sum_{i=1}^{n}a_{i}v_{i}(t)
\]
such that its skewness $\widehat{s},$%
\[
\widehat{s}(a_{1},\ldots,a_{n}):=\left\langle \psi^{3}\right\rangle ,
\]
is maximal. The angular brackets indicate an average over the data set. In
order to get a sensical solution some restriction must be imposed on it. An
appropiate restriction is to fix its variance, say%
\begin{equation}
\left\langle \psi^{2}\right\rangle =1.\label{constr}%
\end{equation}
This choice implies that the dimension of $a_{i}$\ is $\left[  v_{i}\right]
^{-1}.$ In terms of the coefficients $\left\{  a_{i}\right\}  $\ the
constraint reads,%
\begin{equation}
\left\langle \psi^{2}\right\rangle =\sum_{i,j=1}^{n}a_{i}C_{ij}a_{j}%
=1,\label{constC}%
\end{equation}
where $C_{ij}$ are the elements of the covariance matrix $C_{il}:=\left\langle
v_{l}v_{i}\right\rangle .$\ The maximal skewness $\widehat{s}_{m}$ is obtained
by setting equal to zero the variation of $(\widehat{s}-\lambda\left\langle
\psi^{2}\right\rangle )$ where $\lambda$ is the Langrange multiplier
associated with the variance constraint (\ref{constr}). In this way one
obtains%
\begin{equation}
3\sum_{k,l=1}^{n}a_{k}\widehat{S}_{kli}a_{l}-2\lambda\sum_{l=1}^{n}C_{il}%
a_{l}=0,\ \ \ \ \ \ 1\leq i\leq n,\label{eqtn}%
\end{equation}
with
\[
\widehat{S}_{kli}:=\left\langle v_{k}v_{l}v_{i}\right\rangle ,
\]
the elements of the skewness tensor $\widehat{S}.$ We assume all the elements
of the tensors $C$ and $\widehat{S}$ to exist. The values of the
$n$\ coefficients $a_{i}$\ and of the Lagrange multiplier\ $\lambda$
satisfying the $(n+1)$ equations (\ref{constr}) and (\ref{eqtn}) will be
denoted as $a_{i}^{\alpha}$\ and as\ $\lambda_{\alpha}$ respectively. The
number of real solutions maybe larger than $n,$ see for example Fig. \ref{f1}. If
$\left\{  \lambda_{\alpha},a_{i}^{\alpha}\right\}  $\ is a solution then also
$\left\{  -\lambda_{\alpha},-a_{i}^{\alpha}\right\}  $ is a solution. The
value $\lambda_{\alpha}$\ associated with a solution $\overrightarrow
{a^{\alpha}}$ is proportional to the skewness $\widehat{s}(a_{1}^{\alpha
},\ldots,a_{n}^{\alpha})$. This can be seen by multiplying the $i$-th equation
(\ref{eqtn}) by $a_{i}$ summing over all the components and making use of the
constraint (\ref{constC}), one obtains then that%
\begin{equation}
2\lambda_{\alpha}=3\widehat{s}(a_{1}^{\alpha},\ldots,a_{n}^{\alpha
}).\label{lambdaIsSkwns}%
\end{equation}

Without loss of generality, we can take the $\{v_{i}(t)\}$ to be EOFs, i.e.
their covariance matrix $C_{il}:=\left\langle v_{l}v_{i}\right\rangle $ is%
\begin{equation}
C_{il}=w_{i}^{2}\delta_{il}.\label{covarDiago}%
\end{equation}
Then the equations to solve become,%
\[
3\sum_{k,l}a_{k}\widehat{S}_{kli}a_{l}-2\lambda w_{i}^{2}a_{i}%
=0,\ \ \ \ \ \ 1\leq i\leq n.
\]
It is convenient to introduce the dimensionless quantities%
\begin{align*}
\beta_{i}  & :=w_{i}a_{i},\\
\mathrm{and}\ \ \ \ \ \ S_{kli}  & :=\frac{\left\langle v_{k}v_{l}%
v_{i}\right\rangle }{w_{k}w_{l}w_{i}}.
\end{align*}
In terms of these equation (\ref{eqtn}) reads,%
\begin{equation}
3\sum_{k,l}\beta_{k}S_{kli}\beta_{l}-2\lambda\beta_{i}=0,\ \ \ \ \ \ 1\leq
i\leq n,\label{eqtnB}%
\end{equation}
and the unity covariance constraint (\ref{constr}) is,%
\begin{equation}
\sum_{i}^{n}\beta_{i}^{2}=1.\label{unit}%
\end{equation}
A more compact way of writing equation (\ref{eqtnB}) is%
\[
\overrightarrow{\sigma}(\overrightarrow{\beta})=2\lambda\overrightarrow{\beta
},
\]
where $\overrightarrow{\sigma}$ is the gradient of the skewness function
$s(\beta_{1},\ldots,\beta_{n}):=\widehat{s}(a_{1},\ldots,a_{n}),$ i.e.%
\begin{equation}
\sigma_{i}(\overrightarrow{\beta}):=\frac{\partial s}{\partial\beta_{i}}%
=3\sum_{j,k}^{n}S_{ijk}\beta_{j}\beta_{k}.\label{sigma}%
\end{equation}

The solutions to equations ( \ref{eqtnB}) and (\ref{unit}) may correspond to
saddle points of $s(\beta_{1},\ldots,\beta_{n}),$ this is further analyzed at
the end of Section \ref{algorithm}. With $n=2$ the solutions can be expressed
analytically, see the Appendix. With $n>2$ one has to find them numerically,
in the next Section we present a way of doing this.

If all the $w_{i}$'s have the same dimensionality then we can associate
a variance to each solution $\overrightarrow{\beta^{\alpha}},$
namely%
\begin{equation}
W_{\alpha}^{2}:=\sum_{i}^{n}\left(  \beta_{i}^{\alpha}\right)  ^{2}w_{i}%
^{2}.\label{wPeso}%
\end{equation}
It may happen that a solution has a relatively large skewness $s(\beta
_{1}^{\alpha },\ldots ,\beta _{n}^{\alpha })$ while its variance $W_{\alpha
}^{2}$ is relatively small. 
Usually one is interested in solutions with both quantities relatively large.
Accordingly one can consider a dimensional skewness parameter, call it
$\Sigma_{\alpha}$%
\begin{equation}
\Sigma_{\alpha}:=s(\beta_{1}^{\alpha},\ldots,\beta_{n}^{\alpha})W_{\alpha}%
^{3}.\label{dimensionalSkewns}%
\end{equation}

In closing, let us mention that we take the space of the coefficients
$\{\beta_{1},\ldots,\beta_{n}\}$ to be Euclidean, i.e. the inner product of
two $\beta-$vectors is%
\[
\overrightarrow{\beta^{a}}\cdot\overrightarrow{\beta^{b}}=\sum_{i=1}^{n}%
\beta_{i}^{a}\beta_{i}^{b},
\]
and the constraint $\sum\beta_{i}^{2}=1$ means that the vectors
$\overrightarrow{\beta}$ are of length 1.

\section{An algorithm in order to solve the system of equations (\ref{eqtnB})
and (\ref{unit})\label{algorithm}}

Consider a gradient flow, i.e. let $\overrightarrow{\beta}(\tau)$ move
downhill a potential $V(\overrightarrow{\beta}),$
\[
\frac{d\beta_{i}}{d\tau}\mathbf{=}\mathbf{-}\frac{\partial V}{\partial
\beta_{i}},
\]
so that $V(\overrightarrow{\beta})$ is a Lyapunov function,%
\[
\frac{dV}{d\tau}=%
{\displaystyle\sum}
\frac{\partial V}{\partial\beta_{i}}\frac{d\beta_{i}}{d\tau}=-%
{\displaystyle\sum}
\left(  \frac{\partial V}{\partial\beta_{i}}\right)  ^{2}\leq0,
\]
and the evolution of $\overrightarrow{\beta}(\tau)$ stops when an extremum of
$V(\overrightarrow{\beta})$ is reached. In general, such an evolution will not
conserve $\left\vert \overrightarrow{\beta}(\tau)\right\vert ^{2}. $ In this
context notice that if $\left\{  \lambda_{\alpha},\beta_{i}^{\alpha}\right\}
$\ is a solution of (\ref{eqtnB}) then also $\left\{  \mu\lambda_{\alpha}%
,\mu\beta_{i}^{\alpha}\right\}  $\ is a solution of (\ref{eqtnB}). Therefore,
we can work with $\beta$-vectors of arbitrary length if we replace $\lambda$
by $\left\vert \overrightarrow{\beta}\right\vert \lambda,$ i.e. instead of
(\ref{eqtnB}) we can solve%
\begin{equation}
\overrightarrow{\sigma}(\overrightarrow{\beta})=2\lambda\left\vert
\overrightarrow{\beta}\right\vert \overrightarrow{\beta}.\label{fundaScaled}%
\end{equation}
Since both the lhs and the rhs are proportional to $\left\vert \overrightarrow
{\beta}\right\vert ^{2},$ in this formulation the length of $\overrightarrow
{\beta}$ does not play any role. Once a vector satisfying this equation is
found, then we normalize its length in order to obtain a solution of the
equations (\ref{eqtnB}) and (\ref{unit}).

The previous considerations lead us to introduce the following deviation
vector $\overrightarrow{\Delta}(\overrightarrow{\beta}),$%
\[
\overrightarrow{\Delta}(\overrightarrow{\beta}):=\frac{\overrightarrow{\sigma
}(\overrightarrow{\beta})-2\lambda\left\vert \overrightarrow{\beta}\right\vert
\overrightarrow{\beta}}{\left\vert \overrightarrow{\beta}\right\vert ^{2}}.
\]
and a potential $V(\overrightarrow{\beta})$ given by%
\begin{align}
V(\overrightarrow{\beta})  & :=\left\vert \overrightarrow{\Delta}\right\vert
^{2}\label{potential}\\
& =\left\vert \overrightarrow{\beta}\right\vert ^{-4}\left\vert
\overrightarrow{\sigma}(\overrightarrow{\beta})\right\vert ^{2}-12\lambda
\left\vert \overrightarrow{\beta}\right\vert ^{-3}s(\overrightarrow{\beta
})+4\lambda^{2}\geq0.\nonumber
\end{align}
The deviation vector $\overrightarrow{\Delta}(\overrightarrow{\beta})$ does
not depend upon the length of $\overrightarrow{\beta},$ it vanishes when
$\overrightarrow{\beta}$ solves (\ref{fundaScaled}) and $\lambda$ equals the
corresponding Lagrange multiplier $\lambda_{\alpha}.$ Therefore, the value of
$V(\overrightarrow{\beta})$ measures the departure of $\{\lambda
,\overrightarrow{\beta}\}$ from a solution to (\ref{fundaScaled}) and, after
normalization, from a solution to equations (\ref{eqtnB},\ref{unit}). This
specific form of the potential has been chosen because, as it will be seen, it
conserves the length $\left\vert \overrightarrow{\beta}(\tau)\right\vert .$
From it one finds that,%

\begin{equation}
\frac{d\beta_{i}}{d\tau}=\mathbf{-}\frac{\partial V}{\partial\beta_{i}%
}=\left[  \frac{4\left\vert \overrightarrow{\sigma}\right\vert ^{2}%
}{\left\vert \overrightarrow{\beta}\right\vert ^{6}}-\frac{36\lambda
s}{\left\vert \overrightarrow{\beta}\right\vert ^{5}}\right]  \beta_{i}%
+\frac{12\lambda}{\left\vert \overrightarrow{\beta}\right\vert ^{3}}\sigma
_{i}-\frac{12}{\left\vert \overrightarrow{\beta}\right\vert ^{4}}\sum
_{k,j}S_{ijk}\beta_{j}\sigma_{k}.\label{betadot}%
\end{equation}
One can check that indeed this implies $d\left\vert \overrightarrow{\beta
}\right\vert ^{2}/d\tau=0.$ Therefore, we can fix $\left\vert \overrightarrow
{\beta}(\tau)\right\vert ^{2}=1$ and get%
\begin{align}
\left.  \frac{d\beta_{i}}{d\tau}\right\vert _{\left\vert \beta\right\vert =1}
& \mathbf{=}\left[  4\sigma^{2}-36\lambda s\right]  \beta_{i}+12\lambda
\sigma_{i}-12\sum_{kj}S_{ijk}\beta_{j}\sigma_{k}\label{bUnoDot}\\
\mathrm{with}\ \ \ \ \ \overrightarrow{\beta}\cdot\left.  \frac
{d\overrightarrow{\beta}}{d\tau}\right\vert _{\left\vert \beta\right\vert =1}
& \mathbf{=}0\mathbf{.}%
\end{align}

In the definition of the potential $V(\overrightarrow{\beta}),$ equation
(\ref{potential}), and in all the equations we have derived from it, $\lambda$
appears as a free parameter. In order to find the solutions to the equations
(\ref{eqtnB}) and (\ref{unit}) $\lambda$ must take the value\ $\lambda
_{\alpha}$ associated with the vector solution $\overrightarrow{\beta^{\alpha
}}.$\ Therefore one should let also the Lagrange multiplier $\lambda$ evolve
until it reaches the searched value $\lambda_{\alpha}.$ This is achieved by
taking $\lambda(\tau)=\lambda_{M}(\overrightarrow{\beta})$ where $\lambda
_{M}(\overrightarrow{\beta})$ is the $\lambda$\ value that minimizes
$V(\overrightarrow{\beta}),$
\begin{align*}
\left.  \frac{\partial V}{\partial\lambda}\right\vert _{\lambda=\lambda_{M}}
& =0\longleftrightarrow8\lambda_{M}\left\vert \overrightarrow{\beta
}\right\vert ^{3}-12s=0,\\
\mathrm{i.e.}\ \ \ \ \lambda_{M}(\overrightarrow{\beta})  & =\frac{3}%
{2}\left\vert \overrightarrow{\beta}\right\vert ^{-3}s(\overrightarrow{\beta
}).
\end{align*}
Compare this with (\ref{lambdaIsSkwns}). At this value of $\lambda$ the
potential $V(\overrightarrow{\beta})$ equals,%
\begin{align}
V_{M}(\overrightarrow{\beta})  & :=\left.  V(\overrightarrow{\beta
})\right\vert _{\lambda=\lambda_{M}}\label{Vminimum}\\
& =\left\vert \overrightarrow{\beta}\right\vert ^{-4}\left\vert
\overrightarrow{\sigma}(\overrightarrow{\beta})\right\vert ^{2}-9\left\vert
\overrightarrow{\beta}\right\vert ^{-6}s^{2}(\overrightarrow{\beta}%
)\geq0.\nonumber
\end{align}
The evolution of the $\overrightarrow{\beta}(\tau)$ induced by this potential
is,%
\begin{equation}
\frac{d\beta_{i}}{d\tau}\mathbf{=-}\frac{\partial V_{M}}{\partial\beta_{i}%
}=\left[  \frac{4\sigma^{2}}{\left\vert \overrightarrow{\beta}\right\vert
^{6}}-\frac{54s^{2}}{\left\vert \overrightarrow{\beta}\right\vert ^{8}%
}\right]  \beta_{i}+\frac{18s\sigma_{i}}{\left\vert \overrightarrow{\beta
}\right\vert ^{6}}-\frac{12}{\left\vert \overrightarrow{\beta}\right\vert
^{4}}\sum_{kj}S_{ijk}\beta_{j}\sigma_{k}.\label{betaDotMin}%
\end{equation}
In agreement with the results in the previous paragraph, also these equations
lead to $d\left\vert \overrightarrow{\beta}\right\vert ^{2}/d\tau=0.$
Therefore, one could simplify them by taking $\left\vert \overrightarrow
{\beta}(\tau)\right\vert ^{2}=1,$ however, in order to control numerical
errors, it is preferable to use these equations as they stand. As before, one
has that%
\[
\frac{dV_{M}}{d\tau}=%
{\displaystyle\sum}
\frac{\partial V_{M}}{\partial\beta_{i}}\frac{d\beta_{i}}{d\tau}=-%
{\displaystyle\sum}
\left(  \frac{\partial V_{M}}{\partial\beta_{i}}\right)  ^{2}\leq0,
\]
i.e. $V_{M}(\overrightarrow{\beta})$ is the corresponding Lyapunov function.

\begin{figure}[t]
  \noindent\includegraphics[width=0.5\textwidth,angle=-90]{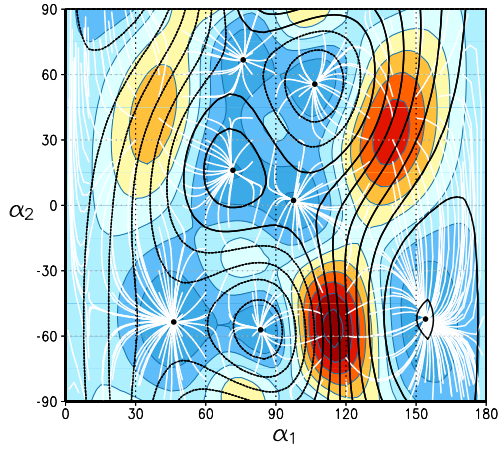}\\
  \caption{The black contours correspond to isolevels of the skewness
$s(\beta_{2},\beta_{7},\beta_{9})$. The shading denotes the values of the potential
$V(\beta_{2},\beta_{7},\beta_{9})$. The black dots indicate the location of the skewness 
maxima, minima and saddles; the non-negative potential vanishes on these locations.
The white lines are the trajectories generated by equation $(\ref{betaDotMin})$ with
initial conditions on a regular lattice.}\label{f1}
\end{figure}

In Fig. \ref{f1} we see an example with $n=3$ constructed from the meteorological
data used in Section (\ref{meteo}). Fig. \ref{f1} shows  the level plots of 
$s(\beta_{2},\beta_{7},\beta_{9})$ and of the
corresponding potential $V(\beta_{2},\beta_{7},\beta_{9})$ on one hemisphere
of the sphere $\beta_{2}^{2}+\beta_{7}^{2}+\beta_{9}^{2}=1.$ The angle 
$\alpha_{1}=\arccos(\beta_{2})$ and the angle $\alpha_{2}=\arctan(\beta_{7}/\beta_{9})$.
The blacks dots
indicate the positions where the potential vanishes, these coincide with the
positions of the two maxima, two minima and three saddles of $s(\beta
_{2},\beta_{7},\beta_{9})$. In addition, Fig. \ref{f1} shows how, starting from
initial values of $\beta_{2},\beta_{7}$ and $\beta_{9},$ on a regular lattice the
gradient-flow equations (\ref{betaDotMin}) lead to these critical values.
Whether a solution $\overrightarrow{\beta^{\alpha}}$ is an extremum or a
saddle depends upon the character of $[h_{ij}^{\alpha}]$ the Hessian matrix at
the solution projected on the sphere $\left\vert \overrightarrow{\beta
}\right\vert ^{2}=1$ , i.e. by the character of the $(n-1)\times(n-1)$ matrix
with elements%
\begin{align*}
h_{ij}^{\alpha}  & :=H_{ij}^{\alpha}-\frac{\beta_{j}^{\alpha}H_{in}^{\alpha
}+\beta_{i}^{\alpha}H_{nj}^{\alpha}}{\beta_{n}^{\alpha}}+\frac{H_{nn}^{\alpha
}\sigma_{i}^{\alpha}\sigma_{j}^{\alpha}-\sigma_{n}^{\alpha}\left[  \beta
_{n}^{\alpha}\delta_{ij}-\beta_{i}^{\alpha}\beta_{j}^{\alpha}\right]
}{\left(  \beta_{n}^{\alpha}\right)  ^{2}},\\
\mathrm{with}\ \ \ \ 1  & \leq i,j\leq n-1,
\end{align*}
where%
\[
H_{li}^{\alpha}:=6\sum_{k}^{n}\beta_{k}^{\alpha}S_{kli}\ \ \ \ 1\leq l,i\leq
n,
\]
is the $n\times n$ Hessian at the solution $\overrightarrow{\beta^{\alpha}} $
and $\sigma_{j}^{\alpha}:=\sigma_{j}(\overrightarrow{\beta^{\alpha}}). $ In
these expressions it is assumed that $\beta_{n}^{\alpha}\neq0.$

\section{Numerical implementation of the gradient algorithm\label{numerical}}

In the general case one has that the number of extrema grows explosively with
$n.$ We deal with this problem by taking first the ten largest EOFs and using
(\ref{betaDotMin}) with 1000 random initial values of 
$\left\{  \beta_{1},\ldots,\beta_{10}\right\}  $ 
in order to find the combination with the
largest skewness, then we increase the number of EOFs by taking the fifteen
largest EOFs and as initial $\beta$-values the solution found in the
previous step suplemented by $\beta_{11}=\cdots=\beta_{15}=0$ and let the 
$\beta$'s evolve according to (\ref{betaDotMin}) with noise added to them. 
Once a
new solution is found the last step is repeated but now with twenty EOFs, etc. The
noise is added in order to explore larger sections of the phase space and not
to remain trapped in a local extremum. A 4th-order Runge-Kutta algorithm is
used in order to integrate the system of ODEs (\ref{betaDotMin}). A solution is found
when the value of the potential $V$ is close enough to zero and the value of the 
skewness has virtually ceased to change.

\begin{figure}[t]
  \noindent\includegraphics[width=0.5\textwidth,angle=-90]{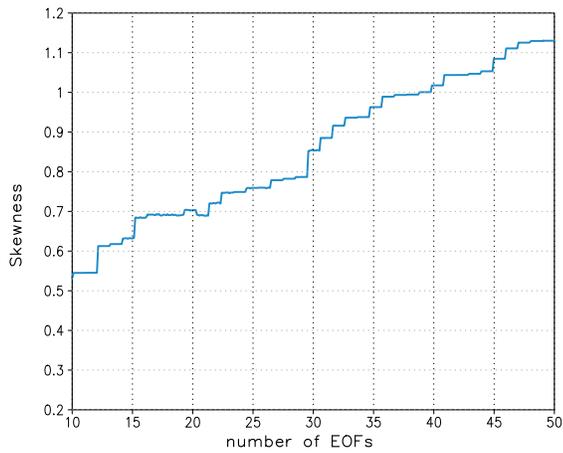}\\
  \caption{Skewness of the maximal-skewness mode as a function of the number of EOFs included
in the calculations.}\label{f2}
\end{figure}

In Fig. \ref{f2} we show the results obtained when this procedure is applied to the
meteorological data used in Section (\ref{meteo}). One can see an increasing
maximal-skewness value as $n$ increases from $10$ to $50.$

\section{Orthogonal set of maximal-skewness modes$\label{ortho}$}

In order to generate a set of linearly uncorrelated orthogonal modes ordered
according to their skewness one should proceed as follows. Firstly, the method
presented in the previous sections is used in order to create from the data
$m(x,t),\ 1\leq x\leq n,$ the linear combination with the largest skewness,
which we write as follows,
\[
\psi^{1}(t)=\sum_{i=1}^{n}\beta_{i}e_{i}(t)=\sum_{i=1}^{n}a_{i}v_{i}(t),
\]
i.e. the $e_{i}(t)$\ are the normalized EOFs,
\[
e_{i}(t)=w_{i}^{-1}v_{i}(t),\ \ \mathrm{with}\ \ \left\langle v_{l}%
v_{i}\right\rangle =w_{i}^{2}\delta_{il}\ \ \ \ 1\leq i,l\leq n.
\]
Thanks to the Karhunen-Lo\`{e}ve theorem \citep{Karhunen1947} 
each $v_{i}$ is associated with a
unique $n$-dimensional eigenvector $\pi_{i}(x)$ satisfying%
\begin{align*}
\sum_{x=1}^{n}C_{yx}^{(m)}\pi_{i}(x)  & =w_{i}^{2}\pi_{i}(y),\ \ \ \ 1\leq
i\leq n,\\
C_{yx}^{(m)}  & =\left\langle m(y,t)m(x,t)\right\rangle ,\ \ \ \ 1\leq x,y\leq
n.
\end{align*}
Therefore, if $w_{i}^{2}\neq w_{j}^{2},$ the corresponding eigenvectors are
orthogonal,%
\[
\sum_{x=1}^{n}\pi_{i}(x)\pi_{j}(x)=\delta_{ij},
\]
and can be taken to be normalized. If the discrete indices $x$ and $y$ are
associated with positions in physical space then each of the eigenvectors
$\pi_{i}(x)$ describes a spatial pattern. In such a case, to $\psi^{1}%
(t)=\sum_{i=1}^{n}a_{i}v_{i}(t)$ there corresponds a unique spatial pattern
$\psi^{1}(x):=\sum_{i=1}^{n}a_{i}\pi_{i}(x).$

Using the covariance metric (\ref{covarDiago}) this $\psi^{1}(t)$ mode is
projected out from the $n$-dimensional set $\left\{  e_{1}(t),e_{2}%
(t),\ldots,e_{n}(t)\right\}  $. The new set so obtained is%
\begin{align*}
\widetilde{m}_{i}(t)  & =(1-\beta_{i}^{2})e_{i}(t)-\beta_{i}\sum_{k\neq
i}\beta_{k}e_{k}\left(  t\right)  ,\\
\left\langle \psi^{1}(t)\widetilde{m}_{i}(t)\right\rangle  & =0,\ \ \ 1\leq
i\leq n.
\end{align*}
Their covariance matrix is%
\begin{align*}
\left\langle \widetilde{m}_{i}^{2}(t)\right\rangle  & =(1-\beta_{i}^{2}),\\
\left\langle \widetilde{m}_{j}(t)\widetilde{m}_{i}(t)\right\rangle  &
=-\beta_{i}\beta_{j},\ \ i\neq j.
\end{align*}
It has one vanishing eigenvalue with eigenvector $\psi^{1}(t)=\sum_{i=1}%
^{n}\beta_{i}e_{i}(t)$ and $(n-1)$-times degenerate eigenvalue 1 with
normalized eigenvectors%
\begin{align*}
\widetilde{e}_{k}(t)  & :=\left(  \beta_{k}^{2}+\beta_{1}^{2}\right)
^{-1}\left[  -\beta_{k}\widetilde{m}_{1}(t)+\beta_{1}\widetilde{m}%
_{k}(t)\right] \\
& =\left(  \beta_{k}^{2}+\beta_{1}^{2}\right)  ^{-1}\left[  -\beta_{k}%
e_{1}(t)+\beta_{1}e_{k}(t)\right]  ,\ \ \ 2\leq k\leq n.
\end{align*}
By construction the $\widetilde{e}_{k}(t)$'s and $\psi^{1}(t)$ are linearly
uncorrelated $\left\langle \psi^{1}(t)\widetilde{e}_{k}(t)\right\rangle =0.$
To each $\widetilde{e}_{k}(t)$ there corresponds a unique pattern
$\widetilde{e}_{k}(x)\propto-a_{k}\pi_{1}(x)+a_{1}\pi_{k}(x).$ It follows that
$\psi^{1}(x)$ and the $(n-1)$ patterns $\widetilde{e}_{k}(x)$ are orthogonal
since%
\begin{align*}
\psi^{1}(x)\widetilde{e}_{k}(x)  & \propto-a_{k}\psi^{1}(x)\pi_{1}%
(x)+a_{1}\psi^{1}(x)\pi_{k}(x)\\
& \rightarrow\sum_{x=1}^{n}\psi^{1}(x)\widetilde{e}_{k}(x)\propto-a_{k}%
a_{1}+a_{1}a_{k}=0.
\end{align*}

As indicated in (\ref{dimensionalSkewns}) instead of $\psi ^{1}(t)$ one may
consider the associated dimensional mode $W_{1}\psi ^{1}(t).$

\section{An application to meteorological data\label{meteo}}

\begin{figure}[t]
  \noindent\includegraphics[width=0.5\textwidth,angle=-90]{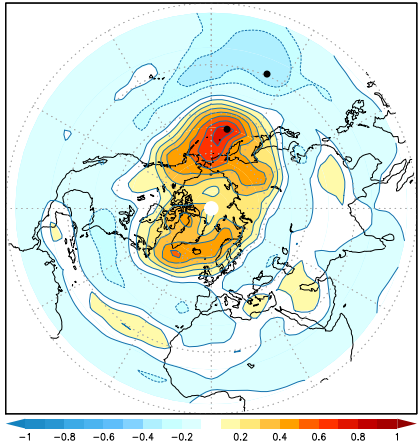}\\
  \caption{Skewness $s(x)$ of the daily streamfunction on the 500 hPa surface in the months
December through February, years 1958 through 2002. The black points indicate the
positions with maximal positive skewness $s(x)=0.79$ and maximal negative skewness
$s(x)= -0.39$.}\label{f3}
\end{figure}

Using the data available at the ECWMF ERA40 website we computed the daily
values of the streamfunction anomalies on the Northern-hemisphere 500 hPa
level during the months December, January and February from 1958 to 2001. 
The skewness in this field was already noticed by \cite{White1980,Nakamuraandwallace1991}.
The dimensionless skewness field $s(x):=\left\langle m^{3}%
(x,t)\right\rangle \left\langle m^{2}(x,t)\right\rangle ^{-3/2}$
 is shown in Fig. \ref{f3}. Two black dots indicate the
two positions with the largest positive and negative skewness $s(x),$ to wit: 
(168 East, 47 North) with skewness 0.76 and (158 East, 19 North) with skewness -0.39.

\begin{figure}[t]
  \noindent\includegraphics[width=0.5\textwidth,angle=-90]{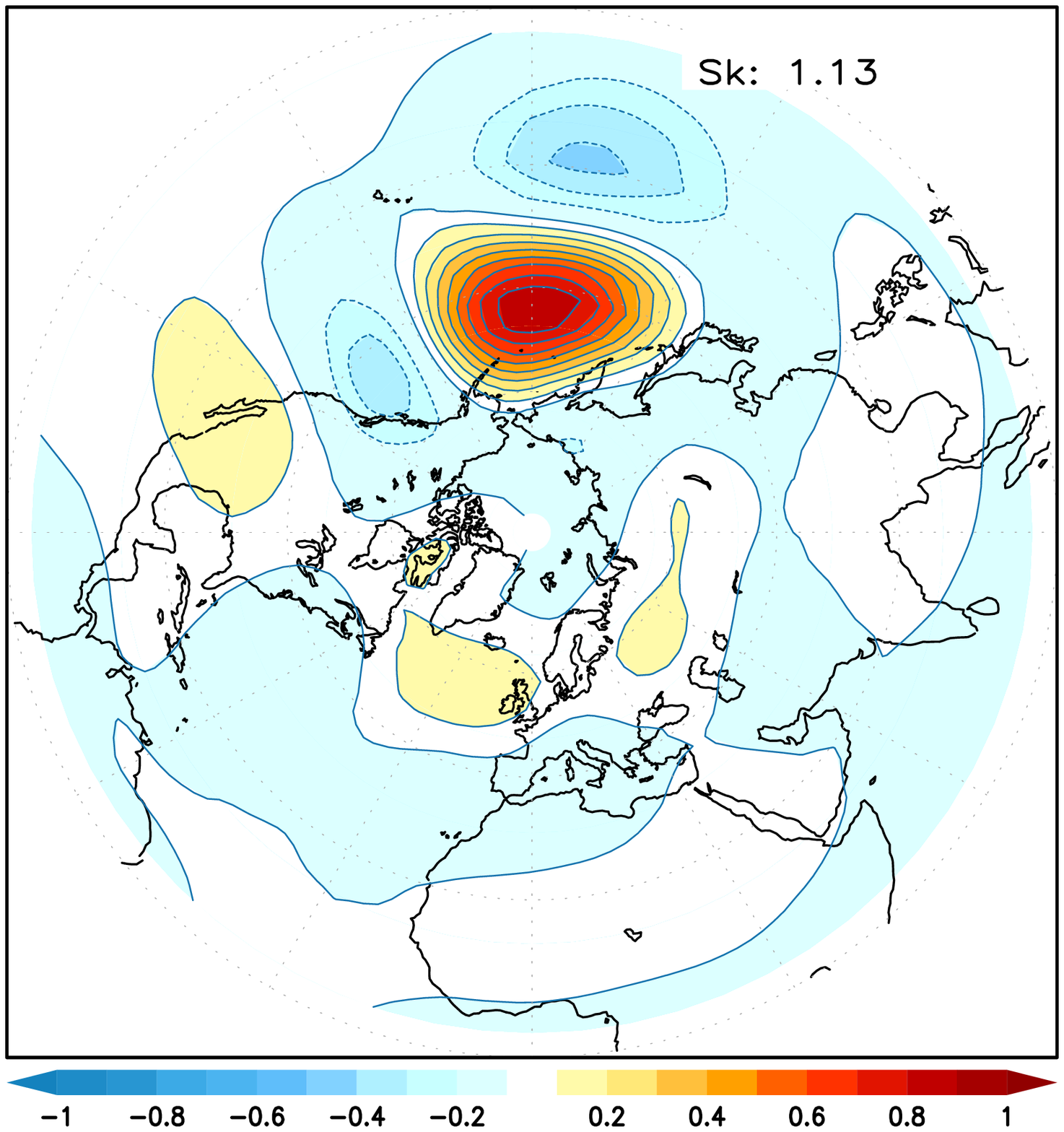}\hfill
  \noindent\includegraphics[width=0.5\textwidth,angle=-90]{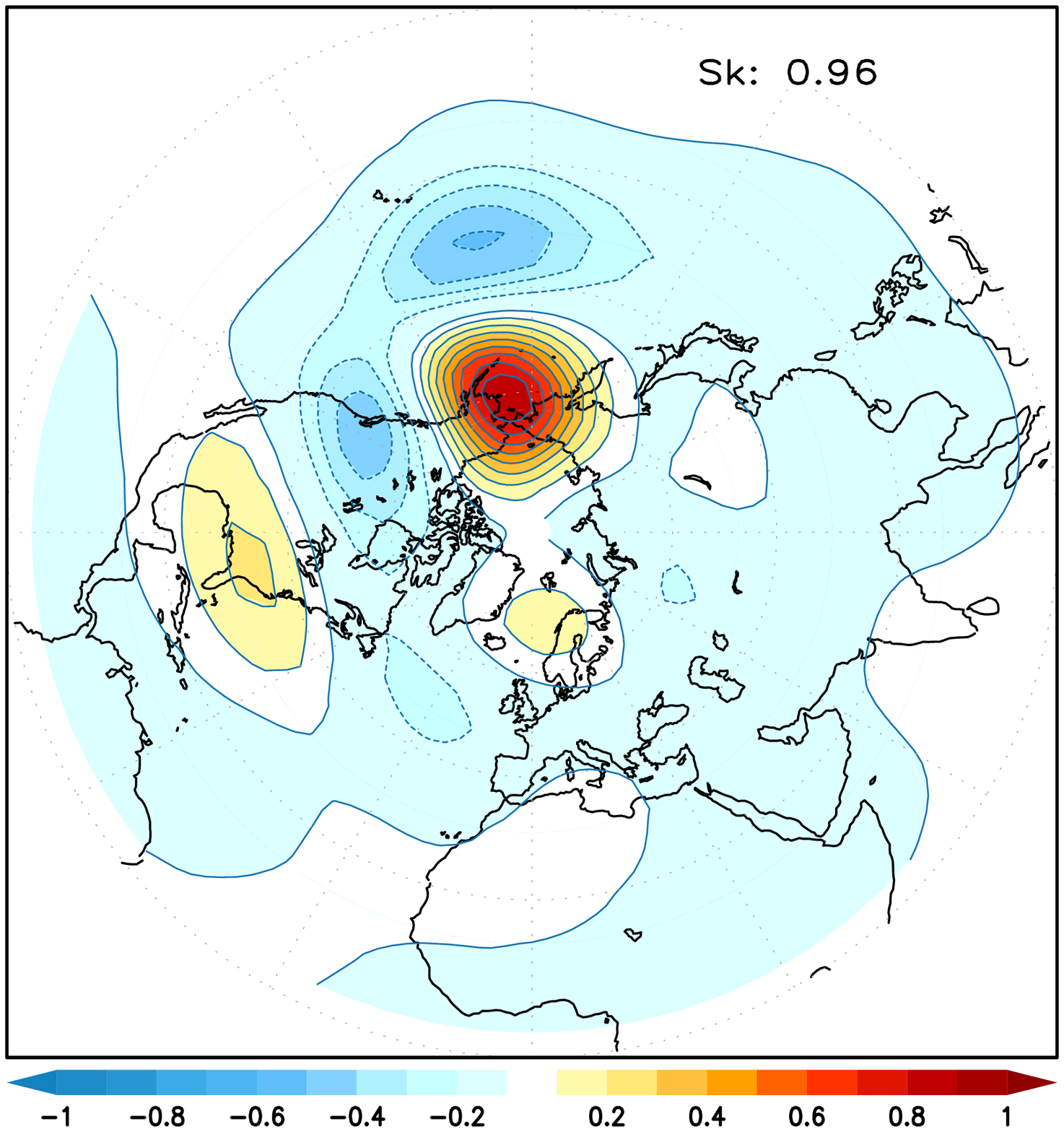}\\
  \caption{The spatial patterns of two maximal-skewness modes of the daily streamfunction on
the 500 hPa surface in the months December through February, years 1958 through
2002. On the left, the largest-skewness mode with skewness
$s_{1} = 1.13$ and, on the right, the mode with next-largest skewness $s_{2} =
0.96$.}\label{f4}
\end{figure}

Fig. \ref{f4} shows the spatial patterns corresponding to $\psi^{1}$ the largest 
and  $\psi^{2}$ the second-largest skewness mode.
The skewness of these modes are $s_{1}= 1.13$ and $s_{2}=0.96 .$ 
That these values are larger than the
maximal values of the local skewness $s(x)$ indicated by the black dots 
in Fig. \ref{f3} is possible 
due to the non-vanishing third-order cross-correlations $\left\langle
m(y,t)m(x,t)m(z,t)\right\rangle \neq0$ for $x,y$ and $z$ not simultaneously identical.

For these data one has that $\sum_{i=1}^{50}w_{i}^{2}\approx 0.9\sum_{i=1}w_{i}^{2},
$\ i.e. the first 50 EOFs describe 90\% of the total variance of the daily
streamfunction fields. In particular, $w_{1}^{2}$ and $w_{2}^{2}$ describe
7.7\% and 6.4\% of the total variance respectively. One also finds that the variances  
$W_{1}^{2}$ and $W_{2}^{2}$ corresponding to the maximal-skewness modes $\psi
^{1}$ and $\psi ^{2},$ confer (\ref{wPeso}), account for 3.3\% and 3.9\% of
the total variance. Their contributions to the total variance is
comparable to those of the EOFs $v_{9}$ and $v_{7}$ respectively. Therefore,
these maximal-skewness modes are physically relevant.

\section{Possible generalizations}

For example, instead of the modes with maximal skewness one could be
interested in the modes with maximal flatness $\widehat{f}(a_{1},\ldots
,a_{n}):=\left\langle \psi^{4}\right\rangle $ given that $\left\langle
\psi^{2}\right\rangle =1.$ The equation analogous to (\ref{eqtnB}) is%
\begin{align*}
4\sum_{k,l,m=1}^{n}\beta_{k}\beta_{l}\beta_{m}F_{klmi}-2\lambda\beta_{i}  &
=0,\ \ \ \ \ \ 1\leq i\leq n,\\
F_{klmi}  & :=\frac{\left\langle v_{k}v_{l}v_{m}v_{i}\right\rangle }%
{w_{k}w_{l}w_{m}w_{i}}.
\end{align*}
The Lagrange multiplier $\lambda$ is proportional now to the flatness of the
solutions, $\lambda_{\alpha}=2\widehat{f}(a_{1}^{\alpha},\ldots,a_{n}^{\alpha
}).$ The corresponding deviation vector and potential are%
\begin{align*}
\overrightarrow{\Delta}_{4}(\overrightarrow{\beta})  & :=\frac{\overrightarrow
{\phi}(\overrightarrow{\beta})-2\lambda\left\vert \overrightarrow{\beta
}\right\vert ^{2}\overrightarrow{\beta}}{\left\vert \overrightarrow{\beta
}\right\vert ^{3}}.\\
V_{4}(\overrightarrow{\beta})  & :=\left\vert \overrightarrow{\Delta}%
_{4}\right\vert ^{2},
\end{align*}
with%
\[
\phi_{i}(\overrightarrow{\beta}):=4\sum_{k,l,m=1}^{n}\beta_{k}\beta_{l}%
\beta_{m}F_{klmi}.
\]
The minimum value of $V_{4}(\overrightarrow{\beta})$ is achieved when
$\lambda$ equals $\lambda_{M}=2\left\vert \overrightarrow{\beta}\right\vert
^{-4}f(\beta_{1},\ldots,\beta_{n}),$ the potential is then%
\begin{align*}
V_{4M}(\overrightarrow{\beta})  & :=\left.  V_{4}(\overrightarrow{\beta
})\right\vert _{\lambda=\lambda_{M}}\\
& =\left\vert \overrightarrow{\beta}\right\vert ^{-6}\left\vert
\overrightarrow{\phi}(\overrightarrow{\beta})\right\vert ^{2}-16\left\vert
\overrightarrow{\beta}\right\vert ^{-8}f^{2}(\beta_{1},\ldots,\beta_{n}).
\end{align*}
As one can check $\sum\beta_{i}\left(  \partial V_{4M}/\partial\beta
_{i}\right)  =0,$ so that $d\left\vert \overrightarrow{\beta}\right\vert
^{2}/d\tau=0.$

\section*{Acknowledgment} 
ECMWF ERA-40 data used in this study is the Basic 2.5 Degree Atmospheric data set in the ECMWF Level III-B archive obtained from their data server at data.ecmwf.int/products/data/archive/.

\section*{Appendix}

With two EOFs and after eliminating $\lambda$ there is only one equation to
solve, namely%
\[
\beta_{2}\left(  \beta_{1}^{2}S_{111}+\beta_{2}^{2}S_{221}+2\beta_{1}\beta
_{2}S_{121}\right)  =\beta_{1}\left(  \beta_{1}^{2}S_{112}+\beta_{2}%
^{2}S_{222}+2\beta_{1}\beta_{2}S_{122}\right)  .
\]
Dividing both sides of this equality by $\beta_{1}^{2}$ one gets that
$z_{2}:=\beta_{2}/\beta_{1}$ is the solution of the following third-order
polynomial,%
\[
z_{2}\left(  S_{111}+z_{2}^{2}S_{221}+2z_{2}S_{121}\right)  =\left(
S_{112}+z_{2}^{2}S_{222}+2z_{2}S_{122}\right)  .
\]
Once $z_{2}$ is known, we recover $\beta_{1}$ and $\beta_{2}$ from%
\begin{align*}
\beta_{1}^{2}+\beta_{2}^{2}  & =1\leftrightarrow\beta_{1}^{2}(1+z_{2}%
^{2})=1\longleftrightarrow\beta_{1}^{2}=(1+z_{2}^{2})^{-1}\\
\ \ \mathrm{and}\ \ \beta_{2}^{2}  & =z_{2}^{2}(1+z_{2}^{2})^{-1}.
\end{align*}


\newpage

\end{document}